\documentclass[epj,final]{svjour}

\usepackage[reqno]{amsmath} \usepackage{latexsym}
\usepackage{graphics} \usepackage{epsfig}

\newcommand{\abs}[1]{\ensuremath{\left|#1\right|}}
\newcommand{\cats}[1]{\ensuremath{\big|#1\big>}}
\newcommand{\bras}[1]{\ensuremath{\big<#1\big|}}
\newcommand{\Proj}{\ensuremath{\mathcal P}}
\newcommand{\UT}{\ensuremath{\mathcal D}}
\newcommand{\sprod}[2]{\ensuremath{\big<#1\big|#2\big>}}
\newcommand{\tr}{\ensuremath{\operatorname{tr}}}

\begin{document}
\title{Excitations of anisotropic spin-1 chains with matrix product
  ground state}\titlerunning{Excitations on MPG-states}
\author{E. Bartel \and A. Schadschneider \and J. Zittartz}
\authorrunning{E. Bartel et al.}
%
%
\institute{ Institut f{\"u}r Theoretische Physik, Universit{\"a}t zu
  K{\"o}ln\\ Z{\"u}lpicher Stra\ss{}e 77, D-50937 K{\"o}ln, Germany }
\date{Received: date / Revised version: date}
%
\abstract{ We investigate a large class of antiferromagnetic spin-1
  chains with nearest neighbour interaction and exactly known matrix
  product ground state.  The spectrum of low-lying excitations is
  calculated numerically by DMRG and exact diagonalisation.
  Spin liquid behaviour with an excitation gap is observed
  meeting the Haldane scenario.
  Further, we compare properties of the
  anisotropic model with the well-known isotropic AKLT model and use
  the analytical single mode approximation to obtain a
  quantitative understanding of the excitation gap. 
  The low-lying excited states can be interpreted in terms of a
  magnon-like elementary excitation.
} 
\maketitle
%

\section{Introduction}
\label{intro}

Quantum spin systems have been studied intensively in recent 
years (for a review, see e.g.\ \cite{Bose00}).
Special interest has been taken in the case of low
dimensions where strong quantum fluctuations destroy magnetic order
\cite{MW66}.  Antiferromagnetic spin chains with $S=1$ have been the
focus of theoretical and experimental studies since
Haldane's conjecture \cite{H83a} about the difference between integer
and half-integer spins.
Antiferromagnetic integer spin chains are predicted to be generically
massive while half-integer spin chains have a vanishing gap which can
be rigorously shown for the isotropic spin-$\frac{1}{2}$ chain using
the Bethe ansatz solution \cite{B31}.  
For the isotropic Heisenberg $S=1$ chain with nearest
neighbour exchange no exact solution has been found, but there is
convincing evidence from numerical studies that a finite gap exists in
the thermodynamic limit \cite{WH93}.

In a most general spin-1 Hamiltonian, bilinear as well as biquadratic
interaction terms along with possible anisotropies emerge.
Consideration of only bilinear exchange with anisotropic interaction
and single-ion anisotropies leads to a well-known two-parameter phase diagram
which includes massive as well as massless phases (see e.g.\ \cite{AS98} 
and references therein).  The spin-1 Hamiltonian with isotropic 
nearest neighbour -- bilinear and biquadratic -- interaction has also 
been focus to extensive theoretical studies, see e.g.\
\cite{Soly87,Pap88,Aff89,FS91,FS93,FS93b,Aff94,FS95,TTI01}.  
This one-parameter model
including the Heisenberg Hamiltonian exhibits a
variety of magnetic phenomena like ferromagnetism, dimerization and a
gapped spin liquid (Haldane) phase.  A
valence-bond-solid model (AKLT model) with exactly known ground state
representing this phase has been introduced by Affleck {\it et al.}
\cite{AKLT87}.  Only a few isotropic models with
antiferromagnetic bilinear and biquadratic exchange possess exactly
known spectra \cite{S75,U76,K89,T82,B82}.  While experiments on many
spin-1 compounds \cite{BMA86,MSBC89,RVR88,DR93,DCB94} suggest that
biquadratic exchange can usually assumed to be small,
Mila and Zhang \cite{MZ00} presented a microscopic model with non-negligible
biquadratic exchange in order to understand a massless
phase of the spin-1 vanadium oxide {LiVGe$_2$O$_6$}.

In \cite{KSZ93} a large class of spin-1 models possessing a unique
matrix product ground state was introduced.  This model class
includes biquadratic terms and anisotropic exchange at the same time
and has been proven rigorously to exhibit exponentially decaying
spin-spin-correlations.  In the isotropic limit it reduces to the 
AKLT model \cite{AKLT87}. While the ground state and some of its
properties have been calculated exactly the excitation spectrum in the
anisotropic case has not been subject of studies yet.  It has only
been known that generically a finite gap is observed,
leading to spin-liquid behaviour.

In this work we present numerical and analytical evidence for a finite
excitation gap and discuss the spectrum  and structure of low-lying 
excitations.
Results from an analytical single mode approximation capture the
nature of the lowest excited states and give an exact upper bound to
the excitation gap.


\section{Model and symmetries}
We consider a spin-1 chain with rotational invariance in the
$(x,y)$-plane, invariance under $S^z \rightarrow -S^z$ and 
translation and parity invariance, i.e.\ invariance under the
exchange $j \leftrightarrow j+1$.  The most general Hamiltonian with
nearest neighbour interactions satisfying the above symmetries
\begin{align}
  \begin{split}
    H = \sum_{j} \Big\{ &J_1 \, \left( S^x_j S^x_{j+1} + S^y_j
      S^y_{j+1} + \Delta S^z_j S^z_{j+1} \right)\\[-2mm]
    + \; &J_2 \, {\left(
        S^x_j S^x_{j+1} + S^y_j S^y_{j+1} \right)}^2 + J_3 \, {\left(
        S^z_j S^z_{j+1} \right)}^2 \\[1mm]
    + \; &J_4 \, \left[ \left(
        S^x_j S^x_{j+1} + S^y_j S^y_{j+1} \right) \left( S^z_j
        S^z_{j+1} \right) + \text{\it h.c.} \right]\\
    + \; &D \left( {S^z_j}^2 + {S^z_{j+1}}^2 \right)+ c\Big\}
  \end{split}\label{e-spinH}
\end{align}
gives rise to a model class with seven parameters including an energy
off-set $c$ and a scale, reducing the number of relevant parameters to
five.  The additive constant $c$ is used to shift the ground state 
energy to $E_0=0$.
The Hamiltonian (\ref{e-spinH}) includes the bilinear, anisotropic 
spin-1 chain ($J_2=J_3=J_4=0$) as well as a
chain with only isotropic bilinear and biquadratic exchange
($\Delta=1,J_2=J_3=J_4,D=0$),
both comprising the Heisenberg model.

A large subclass of the general model class (\ref{e-spinH})
has been found \cite{KSZ93} to possess a unique ground state
which can be written as a product
\begin{equation}\label{eq:g0}
  \cats{\psi_0} = \tr \left( m^{(1)} \cdot m^{(2)} \cdots m^{(L)}
  \right)
\end{equation}
of state matrices
\begin{equation}\label{eq:g}
  m^{(j)} = \left(
\begin{array}{rlrl}
  &\cats{\,0\,}_j & \quad \sqrt{a}&\cats{\!+\!}_j\\[1mm]
  \sqrt{a}&\cats{\!-\!}_j & \sigma&\cats{\,0\,}_j
\end{array}
\right), \quad a \neq 0, \quad \sigma \pm 1
\end{equation}
characterised by the real parameter \footnote{Note that we choose 
the parameter $a$ in a different way compared to \cite{KSZ93}
in order to stress the spin-flip symmetry of the state matrix $m^{(j)}$.}
$a$ and the discrete parameter $\sigma=\pm 1$.
Minimizing not only the total energy,
but the energy of the local two-site Hamiltonian as well,
the matrix product ground state (MPG) belongs to the
class of optimum ground states \cite{KSZ93}.
The parameter $a$ represents a
degree of freedom connected to the anisotropy of the model.
Further, taking excited states into account,
three positive real spectral parameters $\lambda_0, \lambda_1, \lambda_2$
complete the parameter space of the MPG model.
Neglecting a trival scale three relevant
real parameters and one discrete parameter survive.
For models with MPG a convenient representation of the spin-1 
Hamiltonian makes use of spectral parameters and projection operators
(see appendix) rather than spin operators  (\ref{e-spinH}) and shall 
be employed in this paper.

Referring to the spin-1 model class, among others, two unitary transformations
\begin{equation}
\UT^1=\prod_j e^{i\frac{\pi}{2}{S^z_j}^2}, \quad
\UT^2=\prod_{j \in A} e^{i \pi S^z_j},  
\end{equation}
with $A$ denoting the subchain of
even sites,
mapping the parameter space onto itself can be found.  Corresponding
Hamiltonians show identical energy spectra, thus
reducing the parameter space we need to discuss.  The unitary
transformation $\UT^1$
maps the parameter $a$ on the model with $-a$. 
It has no impact on
the momentum.
Therefore, in the following we concentrate on $\abs{a}$.
The second unitary transformation $\UT^2$ changes model parameters 
$(a,\sigma)$ to $(-a,-\sigma)$.
Please note that the momentum of odd-$S^z$ states is modified from $k$
to $\pi-k$ whereas the momentum of states with even $S^z$ remains 
unchanged under this transformation.

While the ground state is determined by only the anisotropy
parameter $a$ and the discrete parameter $\sigma$, examination of the
excitation spectrum requires consideration of the spectral parameters
as well.  Choosing the parameters\footnote{This equals $J_1=\frac{1}{2}, 
\Delta=1, D=0, J_2=J_3=J_4=\frac{1}{6}$.}
$\lambda_0=\lambda_1=\lambda_2=1, a=-2, \sigma=-1$ ,
an isotropic point of the MPG model class is reached.  This special
case corresponds to the AKLT model and has been studied intensively
\cite{AKLT87,AAH88,AKLT88}. Therefore, with fixed
$\lambda_0=\lambda_1=\lambda_2=1$ we vary the parameter $\abs{a}$.

\section{Numerical approach to low-lying excitations}

Low-lying excitations of the model are obtained by using density
matrix renormalisation group (DMRG) \cite{W92,W93} and exact
diagonalisation data.  Our DMRG-program implements the infinite system
algorithm with conservation of the quantum number $S^z$ as well as
periodic boundary conditions.
Though usually a higher accuracy of DMRG data is achieved
for open boundary conditions
we have chosen periodic boundary conditions for a better
comparison with analytical
calculations for spin chains of different lengths.
Moreover, dealing with a gapped system including
a unique matrix product ground state the calculated energies 
are supposed to converge quickly \cite{DMRG}.
This has already been observed for the isotropic case in the vicinity 
of the AKLT point \cite{FS93,FS93b}.
In order to check the accuracy of our
data we tested the convergence of the numerical method related to the
chain length $L$ and to the number $m$ of states kept in each
DMRG-iteration as well.  With $L=100$, $m\leq 120$ reliable gap
information for the MPG model class could be achieved.  Since in real
space DMRG momentum is not preserved we used exact diagonalisation of
smaller chains ($L\leq14$) to discuss the momentum of the excitation
levels.


\subsection{Isotropic interactions (AKLT model)}
\label{subiso}

Before we discuss the general case we briefly recall the results for
the isotropic Hamiltonian ($a=-2$, $\sigma=-1$). 
In fact the excitation spectrum of the AKLT model has been studied using 
numerical and analytical methods \cite{FS93b,Fr93} and is well understood.
The elementary excitations are magnon-type spin-1 particles.
The excitation gap is determined by the lowest energy $E_{S}(k)$ of
the triplet excitations ($S=1$) with
momentum $k=\pi$ which belongs to a discrete branch below a continuum.
The second excited state at $k=\pi$ has energy $3E_1(\pi)$ and spin
$S=3$. On the other hand, the gap at momentum $k=0$ is twice as large
as that at $k=\pi$ and has spin $S=2$ \cite{FS93b}. This suggests that the
corresponding states are scattering states of three resp.\ two
magnons with $S=1$ and momentum $k=\pi$.


\subsection{Influence of anisotropy}
\label{sub_aniso}

\begin{figure}
  \begin{center}
    \epsfig{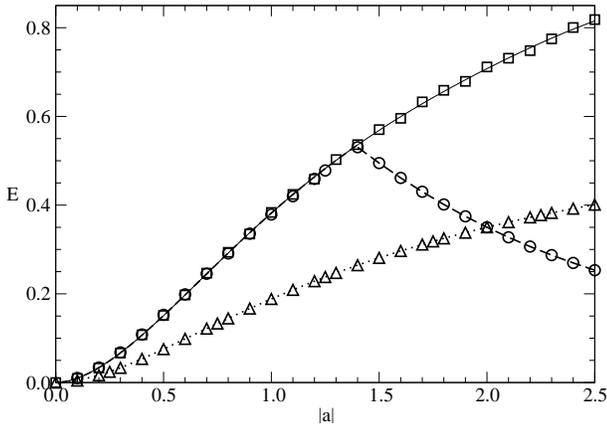}
  \end{center}
\caption{Lowest excitations in the $S^z=0$ ($\circ$), 
$1$ ({\tiny $\triangle$}), $2$ ({\tiny $\Box$}) subspaces plotted 
versus anisotropy parameter $\abs{a}$ (DMRG-data).}
\label{f.fullspectrum}
\end{figure}

In fig.~\ref{f.fullspectrum} the numerically calculated lowest excitations in
three $S^z$-subspaces are plotted as function of the parameter
$\abs{a}$.
For $\abs{a} \neq 0$ a finite excitation gap to the unique ground
state is observed \cite{KSZ93}.
At the point $a=0$ the model reaches a higher symmetry
(invariance under transformation $\UT^1$)
leading to a
three-fold degenerate ground state with vanishing excitation gap
in the thermodynamic limit.
At the isotropic AKLT model with $a=-2$ and $\sigma=-1$ the lowest
excitation is a triplet with total spin $S=1$ and
momentum $k=\pi$, in agreement with previous results
(see Sec.~\ref{subiso}).  A degeneracy of the energy levels in
the $S^z=0,1$ subspaces can thus be seen at $\abs{a}=2$.
For smaller values of $\abs{a}$ the anisotropy is of
easy-plane type leading to a splitting of the lowest excitation level.
Here, excitations along the z-axis
are energetically favourable. Therefore the lowest excitations are
found in the $S^z=1$ subspaces.  In contrast, for $\abs{a}> 2$ 
an easy-axis anisotropy favours excitations with a vanishing
total $S^z$.
Further, we observe a level crossing in the $S^z=0$ subspace at
$a=a_l\approx 1.4$.
An obvious conjecture, consistent with the numerics, would be $a_l=\sqrt{2}$.
At this point the interaction parameter $J_3$ changes sign.
However, for a different choice of anisotropy parameters $\lambda_j$ 
the level crossing seems not to be related to such a change of sign.
It can even be observed in cases where there is no sign change at all
(see Sec. \ref{s.spectral}).

For smaller $\abs{a}$ the lowest excitation with 
vanishing $S^z$
degenerates with the lowest energy levels in the $S^z=2$ subspaces.
Its energy is close to double the size of the $S^z=1$ excitation gap.
This gives rise to the conjecture that the lowest excitations in the
$S^z=0,2$ subspaces consist of two $S^z=1$ excitations.
In the following sections we find more evidence for this picture.
For $a>a_l$ a different energy level in the $S^z=0$ subspace becomes
energetically favourable which, for $a \to \infty$, decreases to zero. 

With DMRG, we calculated up to five lowest eigenenergies of the system
in each $S^z$-subspace.
All higher excitation energies were found to lie,
in the thermodynamical limit, infinitesimally
close to the lowest excitation energy in each subspace
in accordance with an analytical argument of
F\'ath and S\'olyom \cite{FS93}.


\subsection{Dispersion}

\begin{figure*}
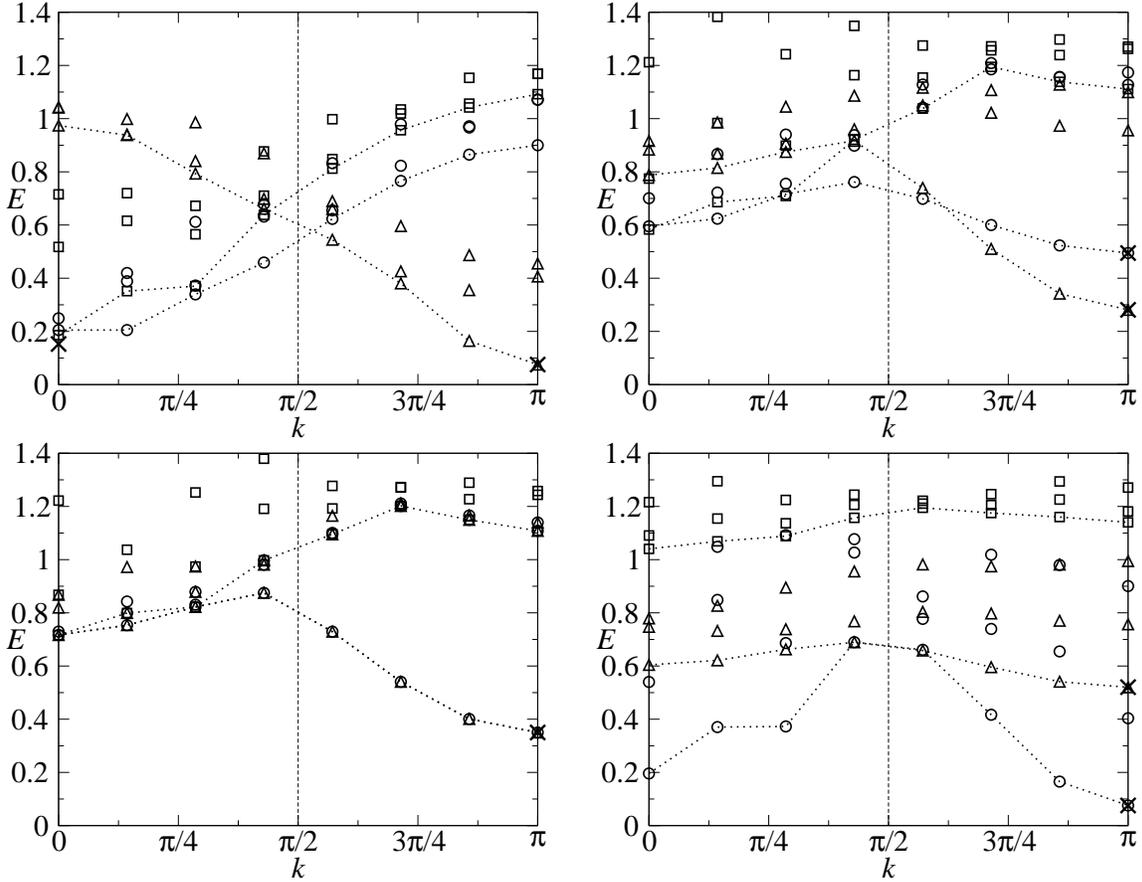

  \begin{center}
    \epsfig{file=PAPER-L14-a0.5-new.eps,height=0.4\linewidth,
      angle=-90, clip=}\qquad
    \epsfig{file=PAPER-L14-a1.5-new.eps,height=0.4\linewidth,
      angle=-90, clip=}\\
    \epsfig{file=PAPER-L14-a2-new.eps,height=0.4\linewidth,
      angle=-90, clip=}\qquad
    \epsfig{file=PAPER-L14-a5-new.eps,height=0.4\linewidth,
      angle=-90, clip=}
  \end{center}
\caption{Dispersion of lowest excitations in the
subspaces for $S^z=0$ ($\circ$),  $S^z=1$ {\tiny $\triangle$} and
$S^z=2$ ({\tiny $\Box$}) at $a=0.5$ (top left), 
$a=1.5$ (top right), $a=2$ (bottom left), $a=5$ (bottom right) 
for $\sigma=-1$. The lowest excitation
energies in each subspace are indicated by a dotted line which only
serves as a guide to the eye. The data  are obtained by exact 
diagonalisation ($L=14$). For comparison the DMRG 
results ($\times$) from Fig.~\ref{f.fullspectrum} have also been included.}
\label{f.l14a}
\end{figure*}

\begin{figure*}
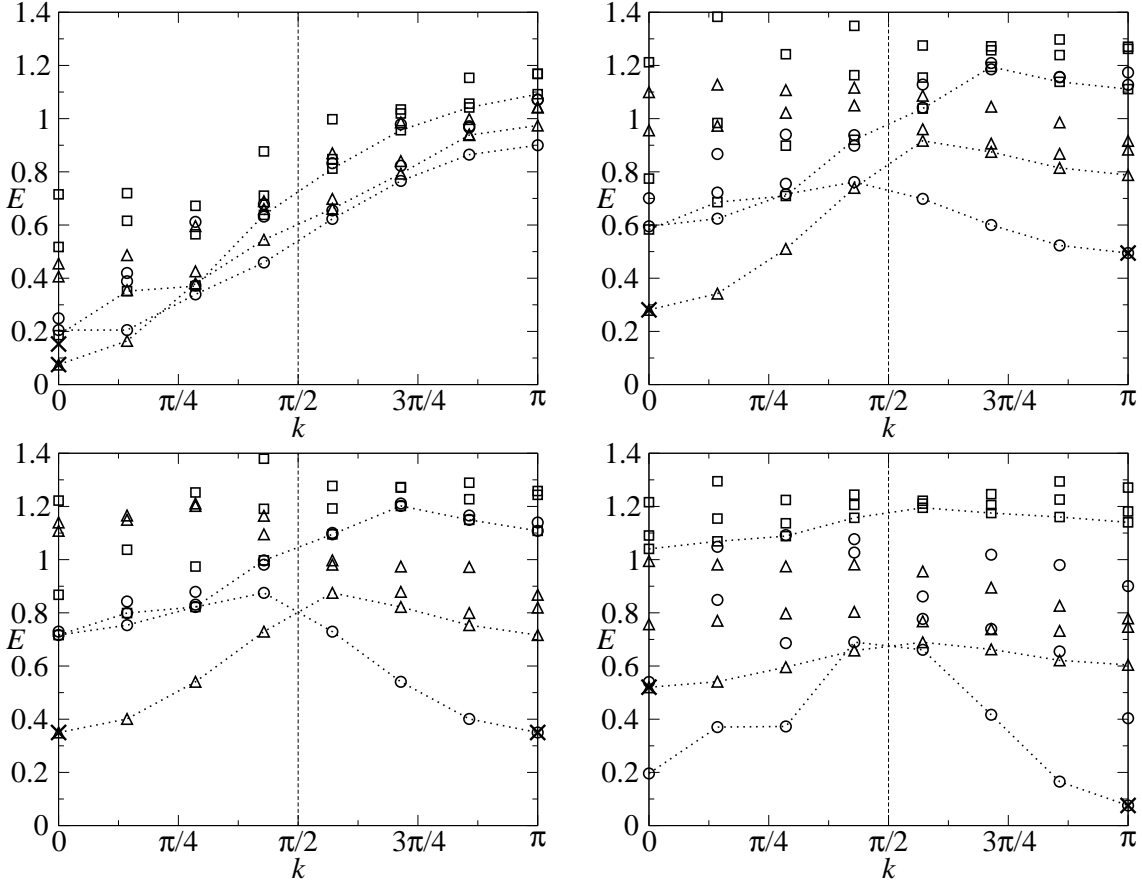

  \begin{center}
    \epsfig{file=PAPER-L14-a0.5.sigma-new.eps,height=0.4\linewidth,
      angle=-90, clip=}\qquad
    \epsfig{file=PAPER-L14-a1.5.sigma-new.eps,height=0.4\linewidth,
      angle=-90, clip=}\\
    \epsfig{file=PAPER-L14-a2.sigma-new.eps,height=0.4\linewidth,
      angle=-90, clip=}\qquad
    \epsfig{file=PAPER-L14-a5.sigma-new.eps,height=0.4\linewidth,
      angle=-90, clip=}
  \end{center}
\caption{Same as Fig.~\ref{f.l14a} with $\sigma=1$.}
\label{f.l14b}
\end{figure*}

As a method in real space, DMRG gives no information about the
momentum of the excitations. In order to have a closer look at the
dispersion curve, the low-energy spectra of smaller chains ($L \leq
14$) were calculated by exact diagonalisation.  
Figures \ref{f.l14a} and \ref{f.l14b} show the dispersion of low-lying 
excitations at four
representative points for $\sigma=-1$ (Fig.~\ref{f.l14a})
and $\sigma=1$ (Fig.~\ref{f.l14b}). 
In order to estimate the accuracy of the exact diagonalization results
we have also included the DMRG data for the energy gaps.
They agree very well with extrapolations from results for systems
with $L\leq 14$. 
Interestingly, large finite-size corrections are only found
in the $S^z=0$ subspace for small $|a|$.
Later we will argue that the corresponding states can be interpreted
as scattering states of two magnons. The other states found using 
DMRG are single magnon excitations. Their energies converge evidently
faster with increasing system size.


In the following subsections we will discuss the spectra for the
regions $0< |a| < a_l$, $a_l< |a| < 2$ and $|a| >2$ separately.
Figs.~\ref{f.l14a} and \ref{f.l14b} show typical examples for each region.


\subsubsection{$0< |a| < a_l$}

For anisotropy parameters $0< |a| < a_l$ the lowest  excitation is found 
in the subspace of $S^z=1$ and has momentum $k=\pi$ ($\sigma=-1$) 
or $k=0$ ($\sigma=1$).
The lowest states in the $S^z=0$ and $S^z=2$ subspaces are degenerate.
They have momentum $k=0$ and an energy which is twice the value of the gap.
The corresponding states might therefore by interpretated as scattering
states of two elementary magnon-type excitations with $S^z=1$.


\subsubsection{$a_l < |a| < 2$}

As already mentioned in Sec.~\ref{sub_aniso} at $a_l\approx 1.4$
a level crossing is observed for the second excited state.
In the region $a_l < |a| < 2$ the scattering states for small momenta
still exist. However, another state with $S^z=0$ has now a lower energy.
The momentum of the corresponding second gap, which is smaller than
twice the gap in the $S^z=1$-subspace, for $\sigma=-1$ is now given 
by $k=\pi$ instead of $k=0$, and vice versa for $\sigma=1$. 
Later (in Sec.~\ref{sub_sz0}) we will show that this new state is 
well described by a single mode approximation and therefore corresponds 
to another single-magnon excitation. This state is result of the splitting of the Haldane triplet due to the anisotropy.


\subsubsection{$ |a|=2$}

For the $|a|=2$, $\sigma=-1$, including the isotropic point for
$a=-2$, the well-known properties 
of the AKLT model are reproduced (Sec.~\ref{subiso}).  
The model's isotropy leads 
to a degeneracy of the $S^z=0,1$ excitation branches indicating the
triplet character of an excitation with total spin~$1$.  As can easily
be seen, the gap still occurs at a momentum $k=\pi$.  

For $|a|=2$ and $\sigma=1$ the model is not isotropic and hence no
degeneracy of the $S^z=0,1$ states is observed. However, the transformation
$\UT^2$ which changes $\sigma\to-\sigma$ also transforms the momenta $k$
of odd-$S^z$ states to $\pi=k$. This leads to the observed symmetry
of the lowest excited states around $k=\frac{\pi}{2}$.


\subsubsection{$ |a|>2$}

For $|a|>2$ the structure of the low-lying excitations changes
compared to the case $|a| < 2$. 
Now the lowest excitation is found in the $S^z=0$ subspace.
We will later show that it again can be interpreted as a magnon.
However, the energy gap still occurs for momentum $k=\pi$.  
Also the second excitation, which now is found in the $S^z=1$ subspace,
takes its minimum energy at $k=\pi$ (for $\sigma=-1$) resp.\ $k=0$ 
(for $\sigma=1$) . 
Note that for a large anisotropy ($|a|=5$ in Figs.~\ref{f.l14a},
~\ref{f.l14b}) all low-lying excitations (except for the first excitated 
state) have almost no dispersion.



\subsection{Variation of the spectral parameters}\label{s.spectral}
\begin{figure}
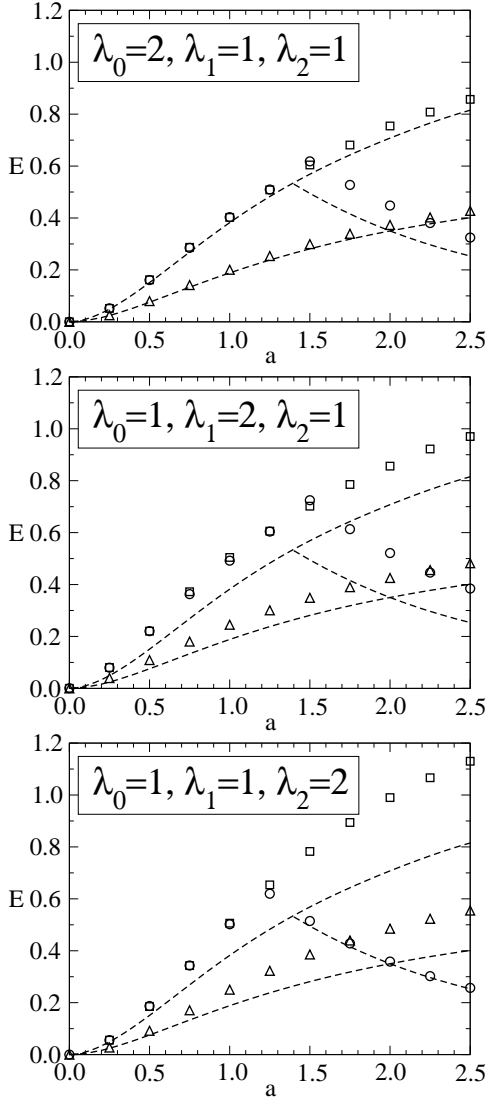

  \begin{center}
    \epsfig{file=lambda-2-1-1.eps,width=0.55\linewidth,
      angle=-90, clip=}
    \epsfig{file=lambda-1-2-1.eps,width=0.55\linewidth,
      angle=-90, clip=}
    \epsfig{file=lambda-1-1-2.eps,width=0.55\linewidth,
      angle=-90, clip=}
  \end{center}
\caption{Variation of the spectral parameters $\lambda_0$, $\lambda_1$, 
$\lambda_2$ ($\circ$ representing the subspace $S^z=0$, {\tiny $\triangle$}  
$S^z=1$ and  {\tiny $\Box$}  $S^z=2$). For comparison: Dashed lines 
represent the original model with $\lambda_0=\lambda_1=\lambda_2=1$. 
The data have been obtained from DMRG.}
\label{f.spectralpars}
\end{figure}

In fig.~\ref{f.spectralpars} DMRG data for models with varying
spectral parameters $\lambda_0, \lambda_1, \lambda_2$ are plotted.
While increasing one of the parameters to $\lambda_i=2$,
the others are kept constant ($\lambda_{j \neq i}=1$).
All diagrams share the qualitative
appearance of the original model with $\lambda_0=\lambda_1=\lambda_2=1$
(fig.~\ref{f.fullspectrum}) containing the isotropic point.
Again, we observe a level
crossing in the $S^z=0$ subspace.
A second level crossing occurs between levels in the subspace with $S^z=0$ 
and $S^z=1$ at the point,
which represents the isotropic system in the original model. 
Note that for the increase of any $\lambda$-parameter 
all excitation levels are lifted compared to the original model.
Yet, the different excitation branches show
different scaling behaviour under the variation of the spectral
parameters.

In the $S^z=0$ subspace with $\abs{a}$ beyond the level crossing we find a 
significant $\lambda_1$-dependence of the decreasing excitation branch.
The influence of the other spectral parameters $\lambda_0$ and
$\lambda_2$ is less important.
The lowest excitations in the $S^z=1$ and $S^z=2$ subspace -- for 
small $\abs{a}$ degenerate with the $S^z=0$ increasing branch -- react to 
variations of the spectral parameters in a similar way.
While $\lambda_1$ has strong influence in the region with small $\abs{a}$
and while increasing $\lambda_2$ has greater impact in the large $\abs{a}$ 
limit, there is only  a weak $\lambda_0$-dependence.  
In models with small $\abs{a}$ states with local spin configuration with
no $z$-axis alignment dominate. Here, we have only a small possibility to find 
$(+,+)$ and $(-,-)$ adjacent spins which contribute to $\lambda_2$-dominated 
excitations. This is different in the large $\abs{a}$ regime where spins 
are strongly aligned along the $z$-axis leading to the observed 
$\lambda_2$-dominance.
 
The similar scaling behaviour of the increasing branch\-es
in the $S^z=0,1,2$ subspaces
suggests a common nature of these excitation levels.
Particularly, it supports the picture that 
the increasing energy branches in the $S^z=0,2$ subspaces are formed
by two weakly interacting magnon-type excitations ($S^z=1$).



\section{Single mode approximation}

In order to substantiate the interpretation of the elementary
excitations as magnons, we use the single mode approximation (SMA).
The SMA is a variational
approach to find approximate low-lying excitations by making use of
ground state correlations.  Therefore, the ground state is locally
perturbed similar to a spin-wave.
This single mode state has a certain momentum
and, since it is orthogonal to the ground state, its energy is an
exact upper bound to the excitation gap. 
Introducing the SMA, Arovas {\it et al.} \cite{AAH88} gave an exact 
upper bound to the excitation gap of the AKLT model. In this paper,
we extend their approach to the general anisotropic MPG model class
in order to illustrate the spectrum of low-lying excitations.

With the locally acting perturbation operator $O_n$ we construct a
translationally invariant state
\begin{equation}
  \cats{\psi_k} = \frac{1}{\sqrt{L}} 
  \sum_{n=1}^L e^{ikn} O_n \cats{\psi_0}
\end{equation}
where $\cats{\psi_0}$ is the exactly known matrix product ground 
state (\ref{eq:g0}).
The energy
\begin{align}
  \begin{split}
  \omega_k
  &= \frac{\bras{\psi_k} H \cats{\psi_k}}
  {\sprod{\psi_k}{\psi_k}}\\
  &= \frac{
    \sum_{n,n'} e^{ik(n-n')} \bras{\psi_0}
    O^\dagger_{n'} H O_{n} \cats{\psi_0}}
    {\sum_{n,n'} e^{ik(n-n')} \bras{\psi_0}
    O^\dagger_{n'} O_{n} \cats{\psi_0}}
  \end{split}
\end{align}
of this variational state
can be regarded as a quotient of  expectation values of the exactly
known matrix product ground state which can be calculated via the
transfer matrix method \cite{KSZ92}. 
It is an exact upper bound to the excitation gap if the
variational state is orthogonal to the ground state.
Complying with conservation of the $z$-component of total spin,
in the following we investigate the subspaces with $S^z=1$ and 
$S^z=0$ separately.
The behaviour of the gap near the boundaries of the MPG region
is of special interest where it is expected to vanish.


\subsection{Subspace $S^z=0$ and behaviour for large $\abs{a}$}
\label{sub_sz0}

Considering a local
application of the $S^z$-operator, the thus built state
\begin{equation}
  \cats{\psi_k^0} =  \frac{1}{\sqrt{L}}
  \sum_{n=1}^L e^{ikn} S^z_n \cats{\psi_0}
\end{equation}
is orthogonal to the ground state $\cats{\psi_0}$ and has a vanishing
total $S^z$-value.
In order to find the lowest excitation with $S^z=0$ the most general
SMA perturbation operator would include contributions of 
$\left({S^z}\right)^2-{a}/{(1+a)}$
which is another locally orthogonal perturbation operator.
However,  a superposition of both possible operator in this subspace
is not mandatory since one can show
that variational states with only $S^z$ as a perturbation operator
already provide the lowest SMA energy levels.

We find the variational
energy
\begin{align}
\begin{split}
  E_k^0 = &\left( \frac{2 \abs{a}}{2 + a^2}\lambda_0 + \lambda_1
  \right)\\ &\left[ \frac{1 + a^2}{( 1+\abs{a} )^2} + \frac{a^2-1}{(
      1+\abs{a} )^2} \cos k \right]
\end{split}  
\end{align}
as an upper bound to the lowest excitation in the $S^z=0$ subspace.

\begin{figure}
  \begin{center}
    \epsfig{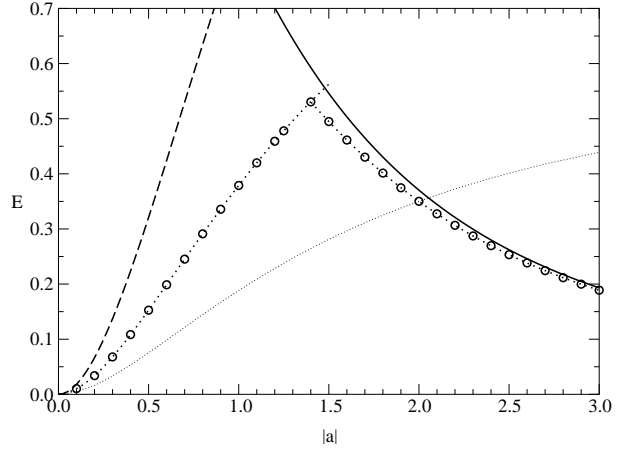}
  \end{center}
\caption{Lowest excitations in the $S^z=0$ subspace. Numerical DMRG data 
(circles) and energy expectation values of the single mode approximation 
(dashed line $k=0$, full line $k=\pi$) are plotted vs. anisotropy 
parameter $\abs{a}$. For comparison the thin dotted line gives the $S^z=1$ 
excitation branch (DMRG data).}
\label{f.smaspectrumsz0}
\end{figure}

In the case of $\abs{a}>2$ the excitation gap is found in the
$S^z=0$ subspace with a momentum $\pi$. The $k=\pi$ variational energy
\begin{equation}\label{e:sma-pi}
  E_\pi^0 = \left( \frac{2 \abs{a}}{2 + a^2}\lambda_0 + \lambda_1
  \right) \frac{2}{( 1+\abs{a} )^2}
\end{equation}
tends to zero for $\abs{a} \to \infty$.
At the AKLT point ($\abs{a}=2$) the SMA variational energy 
$E_{\text{AKLT}}=10/27$ of
Arovas {\it et al.} \cite{AAH88} is reproduced.
For small $\abs{a}$ the lowest $S^z=0$ excited state has a momentum $k=0$.

Figure \ref{f.smaspectrumsz0} shows the SMA energies in comparison to
the excitation gap calculated by DMRG.  For large values of $\abs{a}$ we find
the best agreement between analytical and numerical data in
considering the variational state with momentum $k=\pi$.  While at
the point of the level crossing ($\abs{a} \approx 1.4$) the SMA-energy is
clearly higher than the DMRG data, both analytically and numerically
calculated energies become asymptotically
equal in the $\abs{a} \to \infty$ limit.
For small $\abs{a}$ the lowest excited state of this subspace has vanishing 
momentum.
Here, the variational state with $k=0$ is not capable to replicate
the numerically calculated excitation branch as suitably as in the 
large-$\abs{a}$ case. This indicates a character of the small-$\abs{a}$ 
excitation branch significantly different from the one-magnon picture 
of the excitation gap.


\subsection{Subspace $S^z=1$ and behaviour for small $\abs{a}$}

\begin{figure}
  \begin{center}
    \epsfig{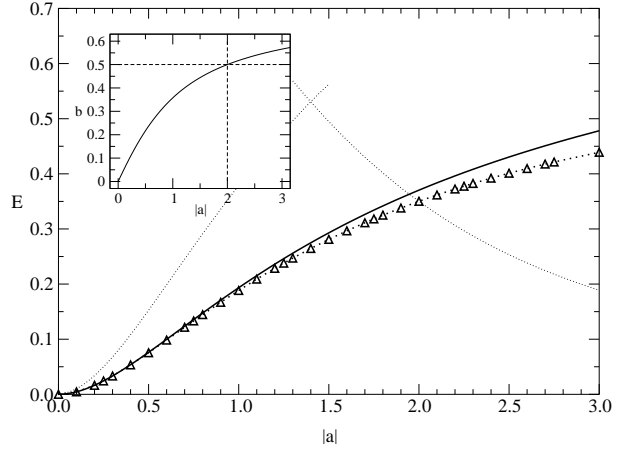}
  \end{center}
\caption{Lowest excitations in the $S^z=1$ subspace. Numerical DMRG data 
({\tiny $\triangle$}) and energy expectation values of the single mode 
approximation (straight line) are plotted vs. anisotropy parameter 
$\abs{a}$. Thin dotted lines give the $S^z=0$ excitation branches 
(DMRG data). The inset shows the minimizing variational parameter 
$b_{\text{min}}$ in the same range of $\abs{a}$.}
\label{f.smaspectrumsz1}
\end{figure}

In order to describe the lowest excitation in the \mbox{$S^z=1$} subspace we
use the most general local perturbation operator
\begin{equation}
  O_n^{(+)}=S^+_n \left[b + (2 b-1)S^z_n\right]
\end{equation}
comprising a variational parameter $b \in {\rm I\!R}$. For $b=0$,
we have $O^{(+)} \propto S^+S^z$ whereas for $b=\frac{1}{2}$ it reduces
to $O^{(+)} \propto S^+$.
The corresponding variational state
\begin{equation}
  \cats{\psi^{+}_k} =  \frac{1}{\sqrt{L}} \sum_{n=1}^L e^{ikn} O_n^{(+)}
\cats{\psi_0},
\end{equation}
is orthogonal to the ground state and has a total $S^z=1$ and momentum $k$.
Since in the MPG model with $\sigma=1$ we expect the excitation gap at 
vanishing momentum, in the following, we set $k=0$ and assume $a>0$ 
without loss of generality.
Square of the norm
\begin{equation}
 \sprod{\psi^{+}}{\psi^{+}}= \frac{1}{1+a}
 \left[{1 + b^2 + \frac{1}{2}a{(1-b)}^2}\right]
\end{equation}
and energy expectation value
\begin{align}
 \begin{split}
 \bras{\psi^{+}}H\cats{\psi^{+}}=& \frac{1}{{(1+a)}^2}
 \left\{
 {\frac{1}{2} \lambda_1 a^2 {(1-b)}^2 + 2 \lambda_2 a b^2} \right.
 \\
 & +\left.
 \frac{2 \lambda_0 a}{2+a^2}{\left[a(1-b)-b\right]}^2 \right\}
 \end{split}
\end{align}
of the variational state can be calculated via the transfer matrix method.
In the following we choose the realistic spectral parameter configuration
$\lambda_0,\lambda_1,\lambda_2=1$, including the AKLT model.
Thus, we obtain the energy
\begin{align}
\begin{split}
  E^{+}&= \frac{\bras{\psi^{+}_k}H\cats{\psi^{+}_k}}
{\sprod{\psi^{+}_k}{\psi^{+}_k}}\\
   &= \frac
      {\frac{1}{2}a^2{(1-b)}^2 + 2 a b^2
      +\frac{2a}{2+a^2}{\left[a-b(1+a)\right]}^2}
      {\left(1+a\right)\left[{1 + b^2 + \frac{1}{2}a{(1-b)}^2}\right]}    
 \end{split}
\end{align}
of the variational state as a function of $a$ and $b$.
Minimizing the energy of the variational state yields the condition
\begin{align}
\begin{split}
  b_{\text{min}} =&
\frac {a^{3} + 4a^{2} + 7a + 6}{a^{2}(a - 2)}\\
&-
\frac {\sqrt{4a^{5} + 36a^{4} + 80a^{3} + 97a^{2} + 84a  + 36}
}{a^{2}(a - 2)}\\
 \end{split}
 \label{e.b}
\end{align}
(inset of  fig.~\ref{f.smaspectrumsz1}).

In fig.~\ref{f.smaspectrumsz1} the energy
of the variational state is depicted.
A good agreement with the numerical data, particularly
for small $a$, is observed.
A series expansion for small $a$
\begin{align}
\begin{split}
  b &= \frac{1}{2} a - \frac{1}{12} a^2
    +\operatorname{O}(a^3),\\
  E^+ &= \frac{1}{2} a^2 - \frac{1}{2} a^3
    + \frac{1}{4} a^4 +\operatorname{O}(a^5),
\end{split}
\end{align}
predicts a decrease of the excitation gap of order $a^2$.
Here, in the $a\to0$ limit, the optimal choice of the
local variational operator is $S^{+}S^z$.
A fit of the numerical DMRG data in this region
\begin{equation}
E_\text{DMRG} \approx 0.500 a^2 - 0.497 a^3 - 0.189 a^4
\end{equation}
can be well reproduced by the variational calculation up to the third
order in $a$.  At the isotropic point ($\abs{a}=2$) we find
$b_{\text{min}}=\frac{1}{2}$ ($S^{+}$ as best variational operator) and
$E_\text{AKLT}=10/27$, both in accordance with the triplet excitation
picture. For large $a$ we find
\begin{align}
\begin{split}
   b_{\text{min}} &= 1 - 2 \sqrt{\frac{1}{a}}
                 + \operatorname{O}(a^{-1}),\\
   E^+ &= 1 - \sqrt{\frac{1}{a}}
                 + \operatorname{O}(a^{-1}).
\end{split}
\end{align}




\section{Conclusion}

The energy gap and lowest excitations of the anisotropic spin-1 chain
with a matrix product ground state have been subject to numerical and
analytical calculations.  DMRG studies showed an excitation gap for
finite values of anisotropy parameter $\abs{a}$. In the isotropic case
the lowest excitation is a triplet of momentum $\pi$. Under
anisotropic interactions this excitation triplet splits and gives rise
to two separate excitation branches with $S^z=0$ and $S^z=1$
establishing the gap for large $\abs{a}$ and small $\abs{a}$,
respectively.

In the large $\abs{a}$ regime the lowest excitation can be identified
as a magnon of vanishing $S^z$ and momentum $k=\pi$.  For $\abs{a}<2$
the excitation gap is found in the $S^z=1$ subspace with momentum
$k=\pi$ ($k=0$) in case of a parameter choice $\sigma=-1$
($\sigma=1$).  Both numerically found excitation energies can be
explained quantitatively well by the single mode approximation.  From DMRG
we find that the $S^z=2$ and the increasing $S^z=0$ excitation
branches have twice the magnitude of the $S^z = 1$ excitation gap.
Together with results from numerical investigation of the momentum and
of the scaling behaviour under variation of the spectral parameters
this is strong evidence that these excitations consists of two weakly
interacting $S^z=1$ magnons.

At $a=0$ a phase transition to a three-fold degenerate ground state with 
excitation continuum occurs. In its neighbourhood we
observed  numerically and analytically
a decrease of the excitation gap with $a^2$.


%
\begin{acknowledgement}
This work has been performed within the research program of
the Sonderforschungsbereich 608 supported by DFG.
\end{acknowledgement}


\section{Appendix A}

Apart from the spin operator description,
the Hamiltonian of the matrix product ground state model
can also be represented by a sum of projection operators on
local ground states.
This spectral representation
reflects the model's symmetries and helps to understand the
free parameters of the MPG model.

Due to the rotational invariance in the $(x,y)$-plane we can divide
the 9-dimensional Hilbert space of a two-site spin-1 system 
into five subspaces of conserved $S^z=S^z_1 + S^z_2$.
The eigenstates of the 1-dimensional subspaces with $S^z=2$
are fixed.
Further, parity invariance leads to discrimination of
symmetric and antisymmetric states.
In the 2-dimensional subspaces characterised by $S^z = 1$
each subspace has one symmetric and one antisymmetric eigenstate.
In case of $S^z=0$, however, after identifying one antisymmetric state,
a 2-dimensional symmetric subspace leaves one degree of freedom
which is described by the superposition parameter $a$.
Finally, we find nine local basis states
\begin{equation}
\renewcommand{\arraystretch}{1.7}
\begin{array}{rlcp{7cm}}
  &\cats{\phi_{\pm2}} &&$ = \cats{\pm \pm}$,\\
  &\cats{\phi_{\pm1}^{\sigma}} &&$ = \frac{1}{\sqrt{2}}
     \left (\cats{\pm 0} +  \sigma \cats{0 \pm}\right)$,\\
  &\cats{\phi_{0}^{-}} &&$ = \frac{1}{\sqrt{2}}
     \left (\cats{+ -} - \cats{- +}\right)$,\\
  &\cats{\phi_{01}^{+}} &&$ = \frac{1}{\sqrt{2a^2+4}}
     \left ( a \left[\cats{+ -} + \cats{- +}\right]  + 2 \cats{0 0}\right)$,\\
  &\cats{\phi_{02}^{+}} &&$ = \frac{1}{\sqrt{2+a^2}}
     \left (\cats{+ -} + \cats{- +} - a \cats{0 0}\right)$
\end{array}\label{e:symbasis2}
\end{equation}
with $\sigma=\pm 1$ distinguishing symmetric and antisymmetric states.
Considering the projection operators 
$\Proj^{\sigma}_{n} := \cats{\phi^{\sigma}_{n}} \bras{\phi^{\sigma}_{n}}$
onto these basis states and paying attention to the spin-flip symmetry
we find a full representation of the local Hamiltonian
\begin{align}
\begin{split}
  h_{j,j+1} &=
      \lambda_2 \left( \Proj_2 + \Proj_{-2} \right)\\
    &+ \lambda_1^{+}  \left( \Proj_1^{+} + \Proj_{-1}^{+} \right)
    + \lambda_1^{-}  \left( \Proj_1^{-} + \Proj_{-1}^{-} \right)\\
    &+ \lambda_{01}^{+} \Proj_{01}^{+} + \lambda_{02}^{+} \Proj_{02}^{+}
    + \lambda_0^{-} \Proj_0^{-} 
\end{split}\label{eq:hspektral}
\end{align}
which depends on six spectral parameters
$\lambda_2$, $\lambda_1^{+}$, $\lambda_1^{-}$,
$\lambda_{01}^{+}$, $\lambda_{02}^{+}$, $\lambda_0^{-}$
and the superposition parameter $a$.
The spin-chain has a MPG
if $\lambda_2, \lambda_1^{-\sigma},\lambda_{02}^{+}>0$
and
$\lambda_1^{\sigma}, \lambda_{01}^{+}, \lambda_0^{-}=0$
leading to a local MPG Hamiltonian
\begin{equation}
\label{eq:hmpg}
  h_{j,j+1} =
      \lambda_2 \left( \Proj_2 + \Proj_{-2} \right)
    + \lambda_1  \left( \Proj_1^{-\sigma} + \Proj_{-1}^{-\sigma} \right)
    + \lambda_{0} \Proj_{02}^{+}
\end{equation}
with $\lambda_1=\lambda_1^{\sigma}$ and $\lambda_0=\lambda_{02}^{+}$.
Spin and spectral parameters of the MPG model class are mapped onto 
each other by 

\begin{align}
  \begin{split}
      J_1 &=
        - \frac{1}{2} \sigma \lambda_{1},\quad
      \Delta =
        - \sigma \frac{\lambda_2}{\lambda_{1}},\quad
      J_2 =
        \frac{1}{2+{a}^{2}}
        \lambda_{0},\\
      J_3 &=
        \frac{a^2}{2+{a}^{2}}
        \lambda_{0}
      - \lambda_{1}
      + \frac{1}{2}\lambda_{2},\\
      J_4 &=
        \frac{a}{2+{a}^{2}}
        \lambda_{0} - \frac{1}{2} \sigma \lambda_{1},\\
      D &=
        \frac{1-a^2}{2+{a}^{2}}
        \lambda_{0}
      + \frac{1}{2} \lambda_{1}, \quad
      c = \frac{a^2-2}{a^2+2}\lambda_{0}.    
  \end{split}
\end{align}



%
%
%
%
\bibliographystyle{phaip.bst}
\bibliography{mpg-paper.bib}

\begin{thebibliography}{10}

\bibitem{Bose00}
I.~Bose,
\newblock Low-dimensional quantum spin systems,
\newblock cond-mat/0011262, 2000.

\bibitem{MW66}
D.~Mermin and H.~Wagner,
\newblock Phys. Rev. Lett. {\bf 17}, 1133 (1966).

\bibitem{H83a}
F.~Haldane,
\newblock Phys. Lett. {\bf 93A}, 464 (1983).

\bibitem{B31}
H.~Bethe,
\newblock Z. Phys. {\bf 71}, 205 (1931).

\bibitem{WH93}
S.~R. White and D.~A. Huse,
\newblock Phys. Rev. B {\bf 48}, 3844 (1993).

\bibitem{AS98}
H.~Aschauer and U.~Schollw{\"o}ck,
\newblock Phys. Rev. B {\bf 58}, 359 (1998).

\bibitem{Soly87}
J.~S\'olyom,
\newblock Phys. Rev. B {\bf 36}, 8642 (1987).

\bibitem{Pap88}
N.~Papanicolaou,
\newblock Nucl.\ Phys.\ B {\bf 305}, 367 (1988).

\bibitem{Aff89}
I.~Affleck,
\newblock J.\ Phys.\ CM {\bf 1}, 3047 (1989).

\bibitem{FS91}
G.~F\'ath and J.~S\'olyom,
\newblock Phys. Rev. B {\bf 44}, 11836 (1991).

\bibitem{FS93}
G.~F\'ath and J.~S\'olyom,
\newblock Phys. Rev. B {\bf 47}, 872 (1993).

\bibitem{FS93b}
G.~F\'ath and J.~S\'olyom,
\newblock J. Phys. CM {\bf 5}, 8983 (1993).

\bibitem{Aff94}
I.~Affleck,
\newblock Rev.\ Math.\ Phys. {\bf 6}, 887 (1994).

\bibitem{FS95}
G.~F\'ath and J.~S\'olyom,
\newblock Phys. Rev. B {\bf 51}, 3620 (1995).

\bibitem{TTI01}
T.~I. K.~Tanaka, A.~Tanaka,
\newblock J.\ Phys.\ A {\bf 34}, 8767 (2001).

\bibitem{AKLT87}
I.~Affleck, T.~Kennedy, E.~H. Lieb, and T.~Tasaki,
\newblock Phys. Rev. Lett. {\bf 59}, 799 (1987).

\bibitem{S75}
B.~Sutherland,
\newblock Phys. Rev. B {\bf 12}, 3795 (1975).

\bibitem{U76}
G.~V. Uimin,
\newblock JETP Lett. {\bf 12}, 225 (1976).

\bibitem{K89}
A.~Kl{\"u}mper,
\newblock Europhys. Lett. {\bf 9}, 815 (1989).

\bibitem{T82}
L.~A. Takhtajan,
\newblock Phys. Lett. {\bf 87A}, 479 (1982).

\bibitem{B82}
H.~M. Babujian,
\newblock Phys. Lett. {\bf 90A}, 479 (1982).

\bibitem{BMA86}
W.~J.~L. Buyers et~al.,
\newblock Phys. Rev. Lett. {\bf 56}, 371 (1986).

\bibitem{MSBC89}
H.~Mutka, J.~L. Soubeyroux, G.~Bourleaux, and P.~Colombet,
\newblock Phys. Rev. B {\bf 39}, 4820 (1989).

\bibitem{RVR88}
J.~P. Renard et~al.,
\newblock J. Appl. Phys. {\bf 63}, 3538 (1988).

\bibitem{DR93}
J.~Darriet and L.~P. Regnault,
\newblock Solid State Commun. {\bf 86}, 409 (1993).

\bibitem{DCB94}
J.~F. DiTusa et~al.,
\newblock Physica B {\bf 194-196}, 181 (1994).

\bibitem{MZ00}
F.~Mila and F.~C. Zhang,
\newblock Eur. Phys. J. B {\bf 16}, 7 (2000).

\bibitem{KSZ93}
A.~Kl{\"u}mper, A.~Schadschneider, and J.~Zittartz,
\newblock Europhys. Lett. {\bf 24}, 293 (1993).

\bibitem{AAH88}
D.~P. Arovas, A.~Auerbach, and F.~D.~M. Haldane,
\newblock Phys. Rev. Lett. {\bf 60}, 531 (1988).

\bibitem{AKLT88}
I.~Affleck, T.~Kennedy, E.~H. Lieb, and T.~Tasaki,
\newblock Commun. Math. Phys. {\bf 115}, 477 (1988).

\bibitem{W92}
S.~R. White,
\newblock Phys. Rev. Lett. {\bf 69}, 2863 (1992).

\bibitem{W93}
S.~R. White,
\newblock Phys. Rev. B {\bf 48}, 10345 (1993).

\bibitem{DMRG}
I.~Peschel, W.~Wang, M.~Kaulke, and K.~Hallberg, editors,
\newblock {\em Density Matrix Renormalisation. A New Numerical Method in
  Physics}, volume 528 of {\em Lecture Notes in Physics},
\newblock Springer, Berlin, 1998.

\bibitem{Fr93}
W.-D. Freitag,
\newblock {\em Anregungen der Spin 1 Valence-Bond-Solid-Kette},
\newblock PhD thesis, Institut f\"ur Theoretische Physik, Universit{\"a}t zu
  K\"oln, 1993.

\bibitem{KSZ92}
A.~Kl{\"u}mper, A.~Schadschneider, and J.~Zittartz,
\newblock Z. Phys. B {\bf 87}, 281 (1992).

\end{thebibliography}
%
%
%

\end{document}